\documentclass[oldversion]{aa}
\usepackage{epsfig,psfig}
\usepackage{txfonts}
\usepackage{color}
\usepackage{graphicx}

\newcommand{\msun}{\mbox{$M_{\odot}$}}

\newcommand{\lsun}{\mbox{$L_{\odot}$}}

\newcommand{\logl}{\mbox{$\log (L/L_{\odot}$)}}  

\newcommand{\zsun}{\mbox{$Z_{\odot}$}}
\newcommand{\teff}{\mbox{$T_{\rm eff}$}}
\newcommand{\Teff}{\mbox{$T_{\rm eff}$}}

\newcommand{\vinf}{\mbox{$\varv_{\infty}$}}
\newcommand{\vesc}{\mbox{$\varv_{\rm esc}$}}

\newcommand{\mdot}{\mbox{$\dot{M}$}}

\newcommand{\msunyr}{\mbox{$M_{\odot} {\rm yr}^{-1}$}}

\def\aap{A\&A}                % Astronomy and Astrophysics
\def\ao{Applied Optics}       % Applied Optics
\def\apj{ApJ}                 % Astrophysical Journal
\def\apjl{ApJ}                % Astrophysical Journal, Letters
\def\apjs{ApJS}               % Astrophysical Journal, Supplement
\def\araa{ARA\&A}             % Annual Review of Astronomy and Astrophysics
\def\mnras{MNRAS}             % Monthly Notices of the RAS
\def\nat{Nature}              % Nature
\def\pasp{PASP}               % Publ. of the Astron. Society of the Pacific
\newcommand{\realcleardoublepage}{\clearpage
  \ifodd \arabic{page}\else \thispagestyle{empty}\mbox{}\newpage \fi }

\newcommand{\be}{\begin{equation}}
\newcommand{\ee}{\end{equation}}

\newcommand{\mstar}{\mbox{$M_{\star}$}}

\newcommand{\kmsec}{\mbox{km\,s$^{-1}$}}

\begin{document}

\title{Fast \& slow winds from supergiants and Luminous Blue Variables}

\author{Jorick S. Vink\inst{1,2}}
\offprints{Jorick S. Vink, jsv@arm.ac.uk}

\institute{$^1$ Armagh Observatory and Planetarium, College Hill, Armagh, BT61 9DG, Northern Ireland, \\
           $^2$ Kavli Institute for Theoretical Physics, University of California, Santa Barbara, CA 93106, USA}

\titlerunning{Fast and slow winds from supergiants and LBVs}
\authorrunning{Jorick S. Vink}

\abstract{We predict quantitative mass-loss rates and terminal wind velocities for early-type supergiants 
and luminous blue variables (LBVs) using a dynamical version of the Monte Carlo radiative transfer method. First, 
the observed drop in terminal wind velocity around spectral type B1 is confirmed by the Monte Carlo method -- 
at the correct effective temperature of about 21\,000 K. This drop in wind velocity is much steeper than would be 
expected from the drop in escape speed for cooler stars. 
The results may be particularly relevant for slow winds inferred for some High-Mass X-ray binaries.
Second, the strength of the mass-loss bi-stability jump 
is found to be significantly larger than previously assumed. Not only could this make bi-stability braking more efficient in 
massive star evolution, but a rotationally-induced version of the bi-stability mechanism may now be capable 
of producing the correct density of outflowing disks around B[e] supergiants, although multi-dimensional modelling including 
the disk velocity structure is still needed. For LBVs, we find the bi-stability jump to become larger 
at higher metallicities, but perhaps surprisingly also larger at {\it lower} Eddington parameters. This may have consequences 
for the role of LBVs in the evolution of massive stars at different metallicities and Cosmic Epochs. Finally, our predicted 
low wind velocities may be important for explaining the slow outflow speeds of supernova type IIb/IIn progenitors, for
which the direct LBV-SN link was first introduced.}


\maketitle

%%%%%%%%%%%%%%%%%%%%%%%%%%%%%%%%%%%%%%%%%%%%%%%%%%%%%%%%%%%%%%%%%%%%%%%%%%%%%%%

\section{Introduction}
\label{s_intro}

Mass loss is an important driver of massive star evolution (Chiosi \& Maeder 1986; 
Langer 2012). This is thought to occur via 
stationary stellar winds on \& off the main sequence (Vink \& Gr\"afener 2012; Groh et al. 2014), and possibly
also in eruptive mode during Luminous Blue Variable (LBV) events 
(Shaviv 2000; Smith \& Owocki 2006; Owocki 2015). Although detailed stationary wind models of LBVs have 
been constructed by Vink \& de Koter (2002) using the Abbott \& Lucy (1985) Monte Carlo
radiative transfer approach, the predicted mass-loss rates (\mdot) were semi-empirical in nature, and assumed 
terminal wind velocities (\vinf). Dynamical modelling has yet to be explored. In this paper,
we will employ the dynamically-consistent approach of M\"uller \& Vink (2008) to predict 
velocity structures and mass-loss rates for a range of OB supergiants and 
LBV models, as well as their metallicity ($Z$) dependence.

Pauldrach \& Puls (1990) first encountered the bi-stability jump 
in modelling the wind of the LBV
 P\,Cygni. Lamers et al. (1995) subsequently found observational evidence  
for the bi-stability jump through a drop in wind velocities by a factor of two 
for a sample of supergiants 
around spectral type B1 -- at an effective temperature of about 21\,000 K (see also Crowther et al. 2006). It 
it was originally assumed that the jump was caused by the optical depth of the Lyman continuum, until
Vink et al. (1999) showed that the recombination of the main
line-driving element iron (Fe) caused an increased amount of line acceleration from 
Fe {\sc iii} -- and an increase in the mass-loss rate by a factor of five. As these 
models were semi-empirical in nature, the 
drop in terminal wind velocity has yet to be theoretically modeled. 

The issue of whether the mass-loss rate increases 
at the bi-stability location, as predicted, or whether it drops instead as suggested 
by empirical results (Trundle et al. 2004; Crowther et al. 2006; 
Benaglia et al. 2007; Markova \& Puls 2008; Morford et al. 2016), remains unresolved, and may depend on 
the question whether the discrepancy may be attributed to macro-clumping, as Petrov et al. (2014) showed 
that the H$\alpha$ line changes its character completely, from an optically thin to an optically thick line below
the bi-stability jump.

In the current paper, we present new dynamically consistent mass-loss predictions on both
sides of the bi-stability jump, and in turns our that we are indeed able to confirm 
the observed drop in terminal wind velocities. Moreover, we predict an even {\it stronger} jump 
in the mass-loss rate by a {\it factor of 10} than the factor of $\sim$5 that 
we found originally, with relevant consequences for 
massive star evolution, including 
the efficiency of bi-stability braking (Vink et al. 2010), the possible formation of B[e] supergiant disks
(Lamers \& Pauldrach 1991), the slow winds in 
High-Mass X-ray binaries (HMXBs), 
LBVs as supernova SN progenitors (Trundle et al. 2008; Groh \& Vink 2011), and 
very massive stars (VMS) as the possible origin of observed 
chemical anti-correlations in globular clusters (Vink 2018).

In Sects.~\ref{s_model} we briefly describe the Monte Carlo modelling and  
physical assumptions. In Sect.~\ref{s_res} mass-loss rates and wind terminal
velocities are presented for a canonical 60 \msun\ supergiant across the temperature regime of the 
bi-stability 
jump, whilst 
Sect.\,\ref{s_lbv} describes similar results for LBVs, characterized by a larger Eddington
$\Gamma$ parameter. The $Z$ dependence is discussed in Sect.\,\ref{s_lbvz}, 
before ending with a summary in Sect.~\ref{s_sum}.

%%%%%%%%%%%%%%%%%%%%%%%%%%%%%%%%%%%%%%%%%%%%%%%%%%%%%%%%%%%%%%%%%%%%%%%%%%%%%%%

\section{Physical assumptions and Monte Carlo modelling}
\label{s_model}

We simultaneously predict mass-loss rates and wind velocity structures 
on the basis of the line-driven wind model of Lucy \& Solomon (1970) and 
Castor, Abbott \& Klein (1975; CAK) 
including multi-line scattering physics (Abbott \& Lucy 1985). 
In the first part of the paper, we expand on the supergiant 
results of Vink et al. (1999), and we subsequently move on to LBV models, 
improving the Vink \& de Koter (2002) results.
The dynamical improvements are based on the M\"uller \& Vink (2008) approach.

The underlying model atmosphere is the Improved Sobolev Approximation code {\sc isa-wind} (de Koter et al. 1993) in which
the effects of the diffuse radiation field are included in the line resonance zones. It  
computes H, He, C, N, O, S, Si, and Fe explicitly in non-LTE, but as we only found minor 
differences when treating Fe in the modified nebular approximation (Schmutz 1991), 
we decided to treat Fe approximately. 
{\sc isa-wind} treats the star (``core'') and wind (``halo'') in a unified manner, i.e. 
there is no core-halo approximation. The temperature is calculated using radiative 
equilibrium in an extended grey LTE atmosphere, and is not allowed to drop below a 
value of half the effective temperature.

In the Monte Carlo part the lines are described in the Sobolev approximation, which 
is an excellent approximation in the outer parts of the winds, where velocity gradients are substantial.
This may provide confidence in our aim of predicting the outer wind dynamics and terminal wind velocity 
correctly.  However, if subtle non-Sobolev effects in the inner wind are relevant, this
may have relevant implications for the predicted values of our mass-loss rates 
(see Krticka \& Kubat 2017). 
Observational and theoretical line transitions have been adopted from Kurucz as previously
(Kurucz \& Bell 1995).

The abundances are taken from Anders \& Grevesse (1989).
Although it has been argued that the overall solar metallicity is smaller than
it was thought to be about a decade ago, the 
Fe abundance does not appear to have changed during this time. Although the effect of a smaller overall 
solar metallicity might lead to lower mass-loss predictions, given the dominance of Fe in setting the mass-loss rate, 
the expected differences may turn out to be relatively small. In any case, the prime reason to keep the 
abundances the same as in previous 
(Vink et al. 1999; 2000; 2001)
predictions is that this approach allows for more straightforward comparisons.

Our 1D wind models are spherically symmetric and homogeneous, although 
wind clumping (micro-clumping) may result in a downward adjustment 
of {\em empirical} mass-loss rates, by a factor 
of $\simeq$3 (Hillier 1991; Moffat \& Robert 1994; Davies et al. 2007; Puls et al. 2008; 
Hamann et al. 2008; Sundqvist et al. 2014; Ram\'irez-Agudelo et al. 2017), whilst it may also affect 
the driving itself (Muijres et al. 2011).

As LBVs find themselves in close proximity to the observed Humphreys-Davidson limit, which is thought 
to be associated with the theoretical Eddington limit, additional physics may lead to 
the development of porous structures (van Marle et al. 2008; Gr\"afener et al. 2012;
Jiang et al. 2015). Porosity effects on mass-loss predictions 
were also investigated by Muijres et al. (2011), where it was noted that it is unlikely that 
predictions would change dramatically. However, porosity (or macro-clumping) may have 
important implications on observational indicators (Oskinova et al. 2007; Surlan et al. 2013; 
Sunqvist et al. 2014; Petrov et al. 2014).

\section{A dynamically consistent bi-stability jump}
\label{s_res}

\begin{figure}
\centerline{\psfig{file=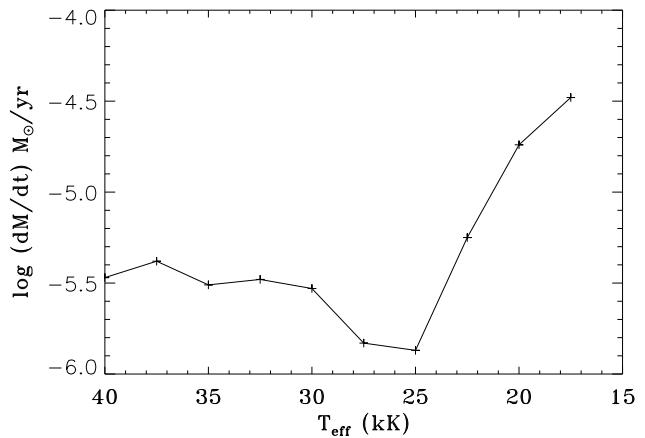, width = 9 cm}}
\caption{The predicted mass-loss rates ($\mdot$) versus \Teff\ for a 60 \msun\ model.}
\label{f_mdot}
\end{figure}

\begin{figure}
\centerline{\psfig{file=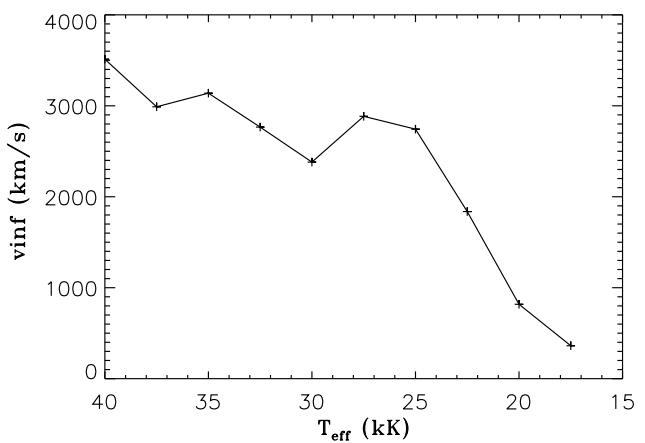, width = 9 cm}}
\caption{The predicted wind velocity ($\vinf$) versus \teff\ for a 60 \msun\ model.}
\label{f_vinf}
\end{figure}

\begin{figure}
\centerline{\psfig{file=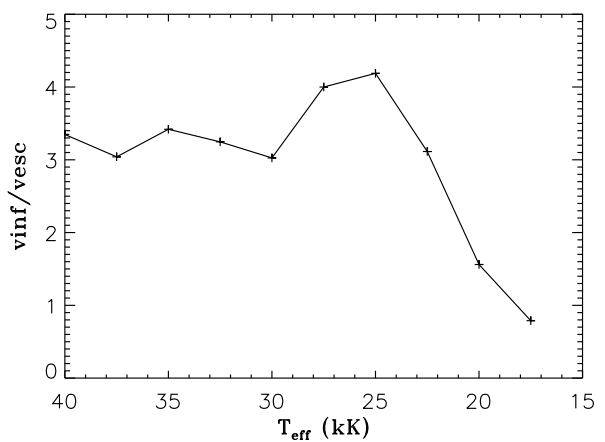, width = 9 cm}}
\caption{The predicted ratio between the wind velocity ($\vinf$) and the escape velocity (\vesc) versus \teff\ for the
60 \msun\ model.}
\label{f_ratio}
\end{figure}

\begin{figure}
\centerline{\psfig{file=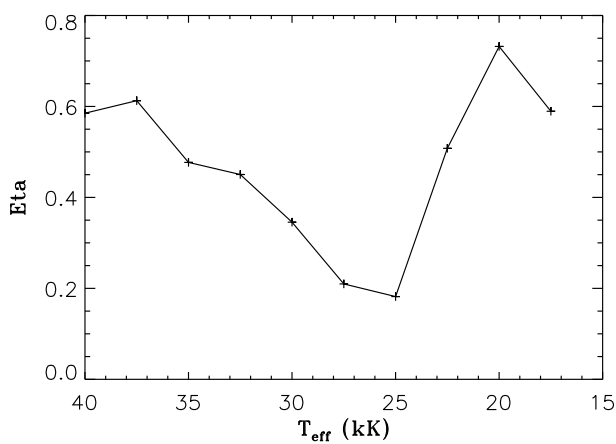, width = 9 cm}}
\caption{The predicted wind efficiency ratio $\eta$ versus \teff\ for the
60 \msun\ model.}
\label{f_eta}
\end{figure}

\begin{table*}
\centering
\begin{tabular}{llll|l|ccc}
\hline
\hline\\[-6pt]
\mstar  & $\log L$ & \teff & $Z/\zsun$ & \vesc    & \vinf   & $\log \mdot$ & \(\beta\) \\[2pt]
[\msun] & [\lsun] &  (kK)  &           & [\kmsec] & [\kmsec]& [\msunyr]   &         \\[3pt]
\hline\\
60      & 6.0     & 40       & 1.0       & 1049  &  3512    & $-$5.47     & 0.99\\
        &         & 37.5     &           & 983   &  2989    & $-$5.38     & 1.02\\
        &         & 35       &           & 918   &  3139    & $-$5.51     & 1.14\\
        &         & 32.5     &           & 852   &  2766    & $-$5.48     & 1.10\\
        &         & 30       &           & 787   &  2381    & $-$5.53     & 1.08\\
        &         & 27.5     &           & 721   &  2884    & $-$5.83     & 1.28\\
        &         & 25       &           & 655   &  2743    & $-$5.87     & 1.40\\
        &         & 22.5     &           & 590   &  1837    & $-$5.25     & 1.10\\
        &         & 20       &           & 524   &  818     & $-$4.74     & 0.81\\
        &         & 17.5     &           & 459   &  362     & $-$4.48     & 0.69\\
\hline\\[-7pt]
\hline
\end{tabular}
\caption{Wind predictions for a 60 \msun\ model with stellar parameters identical
to model series \#10 from Vink et al. (2000).}
\label{tab:results}
\end{table*}

Table~\ref{tab:results} lists the Monte Carlo predictions for our canonical
60 \msun\ supergiant over a range of effective temperatures. 
The stellar parameters are identical to model series \#10 of Vink et al. (2000). 
They are listed in columns (1) - (5), whilst 
predicted wind terminal velocities, new mass-loss rates, and 
wind acceleration parameter $\beta$ from $v(r) = \vinf (1 - r/R)^\beta$ are 
listed in columns (6) - (8). 
The mass-loss predictions are also shown in Fig.~\ref{f_mdot}, and 
the predicted terminal wind velocities are displayed in Fig.~\ref{f_vinf}.

Figure 1 shows that the mass-loss rates increase dramatically {\it by an order of magnitude} between  
25\,000 and 20\,000 K. Moreover, the terminal wind velocity is found to drop significantly over the same 
temperature range (Fig.\,2). This second result is new, whilst the first result is more dramatic
than the factor of five found from the semi-empirical Vink et al. (1999) models, although the 
overall mass-loss rates are in reasonable agreement with the Vink et al. (1999; 2000; 2001) rates.

Figure 2 shows relatively low terminal wind velocities on the cool side of the bi-stability jump 
down to 400 km/s, instead of values in the range 2000-3500 km/s for hotter objects. As the stellar 
escape velocity also drops at lower \teff\ due to the larger stellar radii, it is more insightful
to consider the ratio \vinf\ over \vesc\ instead. This ratio is plotted in Fig.\,3. It is seen 
that the ratio drops rather steeply from values larger than 3 at hot temperatures to values below 1 
on the cool side of the bi-stability jump. 

Note that observationally, Crowther et al. (2006) found 
somewhat smaller values for this ratio with an average value of 3.4 on the hot side of the bi-stability range, 
and somewhat higher values on the cool side (an average value of 1.9). This raises the question whether 
our predicted wind velocities are too high on the hot side of the bi-stability jump, and too low
on the cool side, which would have consequences for our predicted mass-loss rates as well. 
For this reason it is helpful to consider the wind momentum efficiency number $\eta$ which is defined 
as the ratio between the wind momentum per unit time ($\dot{M} \vinf$) over the momentum of the 
radiation field per unit time ($L/c$), or 
$\eta = \dot{M} \vinf / (L/c)$, and displayed in Fig.\,\ref{f_eta}. Similar to the earlier computations of 
Vink et al. (2000), the $\eta$ behaviour shows an increase in the wind efficiency by a factor 2-3, from 25\,000 K onwards, 
now peaking at 20,000 K. This bi-stability temperature
is in agreement with the peak temperature of our alternative {\sc cmfgen} approach of 
Petrov et al. (2016), as well as the observed bi-stability location around spectral type B1 (Lamers et al. 
1995; Crowther et al. 2006).

As the mass-loss rate discrepancy is considered to be unresolvable at the current time until 
appropriate atmosphere modelling including macro-clumping becomes a reality, the issue of the 
wind terminal velocity deserves extra attention, as the terminal wind velocity is generally 
considered to be the more robust empirical wind parameter. Their values are generally
derived from the maximum blue-shifted absorption in resonance lines in the ultraviolet 
part of the spectrum. Systematic errors may work in both directions, and it may be 
difficult to see how empirical values could be underestimated on the hot side 
of the bi-stability jump, whilst being overestimated on the cool side of the jump.

However, it is not inconceivable that this would indeed be the case when 
the wind physics completely changes at 21\,000 K. It is for instance well 
possible that for the faster winds on the hot side of the bi-stability 
jump the measured lines have not yet reached their predicted terminal 
wind speeds yet, resulting in an underestimation of \vinf, whilst the slower 
winds on the cool side of the jump may be overestimated due to an increased wind 
turbulence. This is of course rather speculative, and this would require further 
investigation to find out whether 
the outflow speeds on the cool side of the jump are indeed as fast as 
derived empirically, or as slow as predicted by our Monte Carlo modelling.

A stronger bi-stability jump may have important consequences for 
bi-stability braking (Vink et al. 2010; Markova et al. 2014; Keszthelyi et al. 2017) as well as 
the formation of dense disks 
around B[e] supergiants via the 
rotationally-induced bi-stability jump mechanism of Lamers \& Pauldrach (1991). The basic idea of this model
is that the cooler stellar equator has a higher mass flux and lower wind velocity than the hotter pole, due to the 
Von Zeipel effect. In the updated
computations of Pelupessy et al. (2000) that employed the wind parameters of Vink et al. (1999) this resulted 
in a density contrast between the equator and the pole of a factor of 10. At the time this was deemed insufficient
to explain the dense disks of B[e] supergiants, which lead Cur\'e et al. (2005) to combine the bi-stability mechanism
with slow wind solutions at very high rotation speeds. 

Our new bi-stability models here predict 
both an order of magnitude increase in the mass-loss rate, as well as an order of magnitude drop in the wind velocity.
In the 2D models of Pelupessy et al. (2000) and M\"uller \& Vink (2014) this implies a density contrast between
the stellar equator and the pole by a factor of 100, which may be sufficient
to explain the disk density of B[e] supergiants (Kraus 2017). 
Future multi-dimensional computations are needed to find out if the disk can remain this high density in 
the presence of non-radial line forces (Owocki et al. 1996), whilst we also require an explanation for the 
outflow speeds of just tens of km/s (Kraus et al. 2010; Cidale et al. 2012; Kraus et al. 2016).

Another interesting puzzle regarding wind velocities in massive stars involves the obscured supergiant HMXB 
IGR J17252-3616
uncovered by {\sc Integral}. Manousakis et al. (2012) performed hydro-dynamical modelling, which 
appears similar to the unobscured classical system Vela X-1, but the authors could only explain the obscured 
system with a slow wind of order 500 km/s. 
A possible explanation for the low terminal velocity in IGR J17252-3616 and other HMXBs, such as Vela X1 (see Sander et al. 
2018) would be that the supergiant donor star 
might be located on the cool side of the bi-stability jump where the outflow velocity is found 
to be lower. However, more work is needed to study the ionization and wind physics in the donor stars of 
HMXBs.

\section{Mass loss from Luminous Blue Variables}
\label{s_lbv}

For the second part of the paper we extend the supergiant computations to 
models characteristic for LBVs. The defining property of LBVs is their S Doradus 
variability over timescales of years (Humphreys \& Davidson 1994; Vink 2012), which
might be explained by stellar radius inflation (Gr\"afener et al. 2012). This property enables individual stars 
to cross the bi-stability jump on relatively short evolutionary timescales.

\begin{figure}
\centerline{\psfig{file=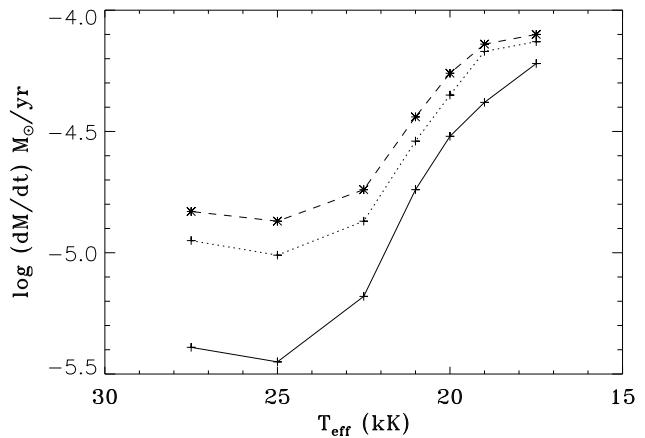, width = 9 cm}}
\caption{Predicted mass-loss rates ($\mdot$) versus \Teff\ for 3 different mass \& Eddington $\Gamma$ values. The 
solid line is for $M = 35\,\msun$, the dotted line for $M = 25\,\msun$, whilst the dashed line is for $M = 23\,\msun$, 
representing $\Gamma$ values of respectively 0.5, 0.7 and 0.8.}
\label{f_lbv-mdot}
\end{figure}

\begin{figure}
\centerline{\psfig{file=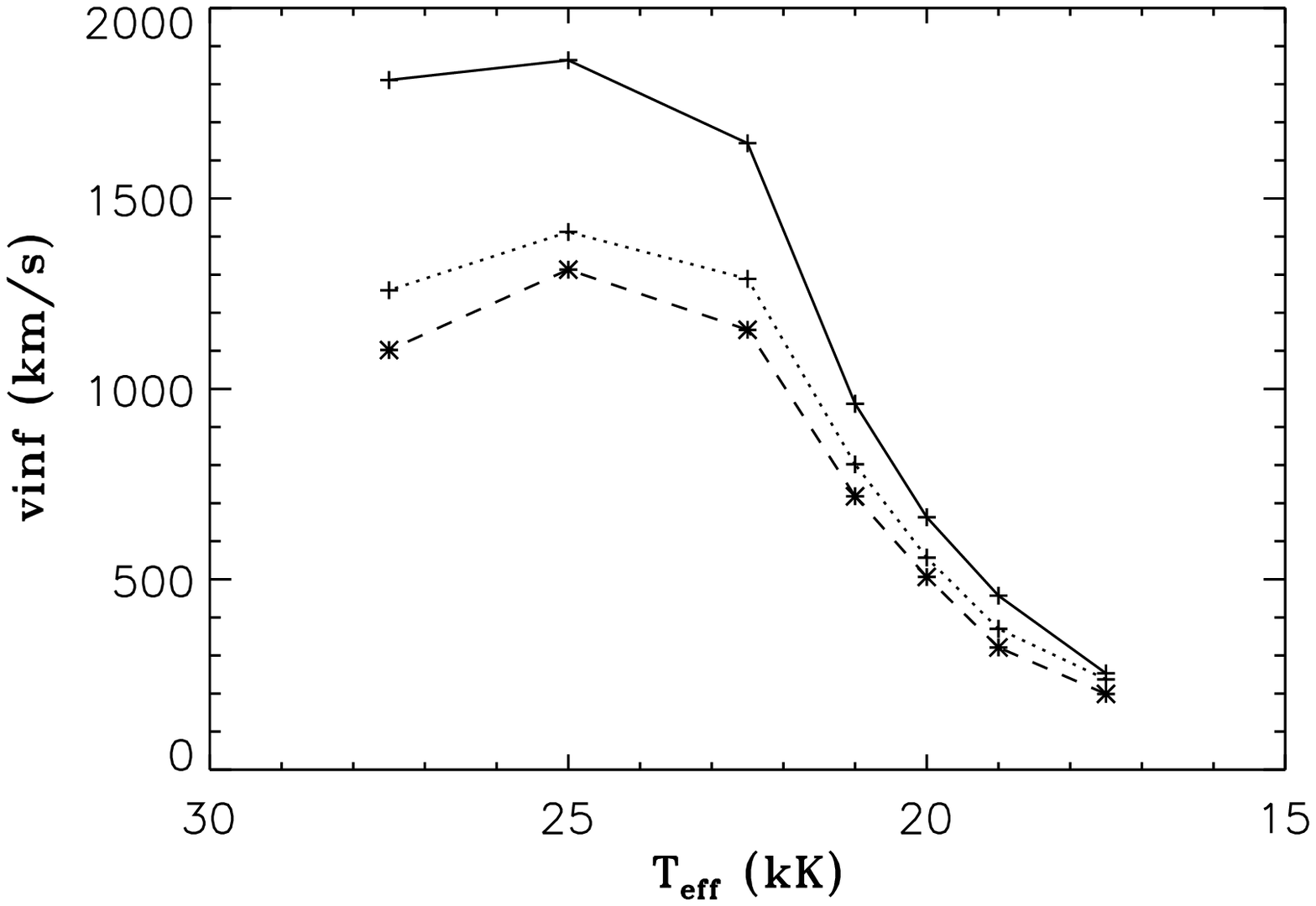, width = 9 cm}}
\caption{The predicted wind velocity ($\vinf$) versus \teff\ for the 3 different LBV mass models. The
solid line is for $M = 35\,\msun$, the dotted line for $M = 25\,\msun$, whilst the dashed line is for $M = 23\,\msun$,
representing $\Gamma$ values of respectively 0.5, 0.7 and 0.8.}
\label{f_lbv-vinf}
\end{figure}

\begin{figure}
\centerline{\psfig{file=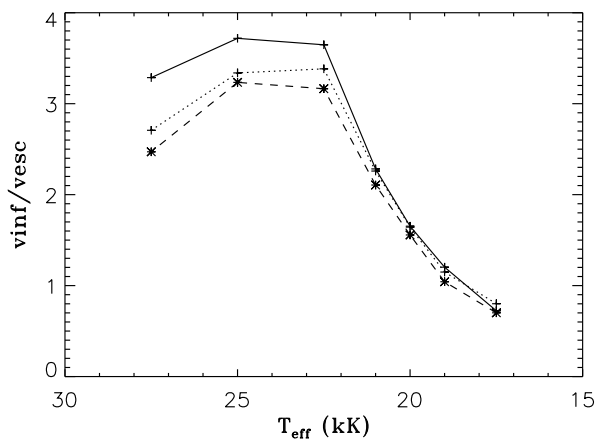, width = 9 cm}}
\caption{The predicted ratio between the wind velocity ($\vinf$) and the escape velocity (\vesc) for the 3 different
LBV mass models. The
solid line is for $M = 35\,\msun$, the dotted line for $M = 25\,\msun$, whilst the dashed line is for $M = 23\,\msun$,
representing $\Gamma$ values of respectively 0.5, 0.7 and 0.8.}
\label{f_lbv-ratio}
\end{figure}

\begin{figure}
\centerline{\psfig{file=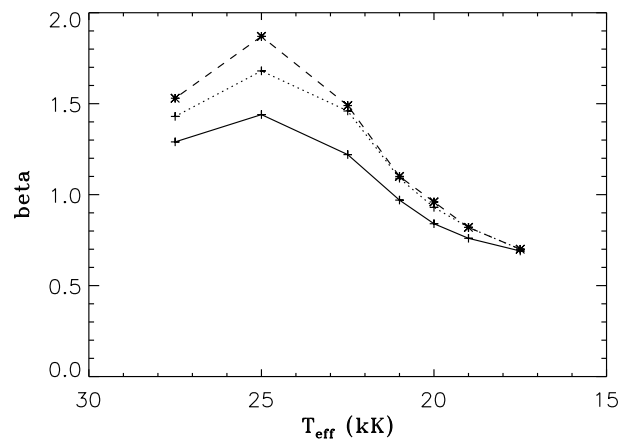, width = 9 cm}}
\caption{The predicted wind structure parameter $\beta$ versus \teff\ for the 3 different LBV mass models. The
solid line is for $M = 35\,\msun$, the dotted line for $M = 25\,\msun$, whilst the dashed line is for $M = 23\,\msun$,
representing $\Gamma$ values of respectively 0.5, 0.7 and 0.8.}
\label{f_lbv-beta}
\end{figure}

\begin{figure}
\centerline{\psfig{file=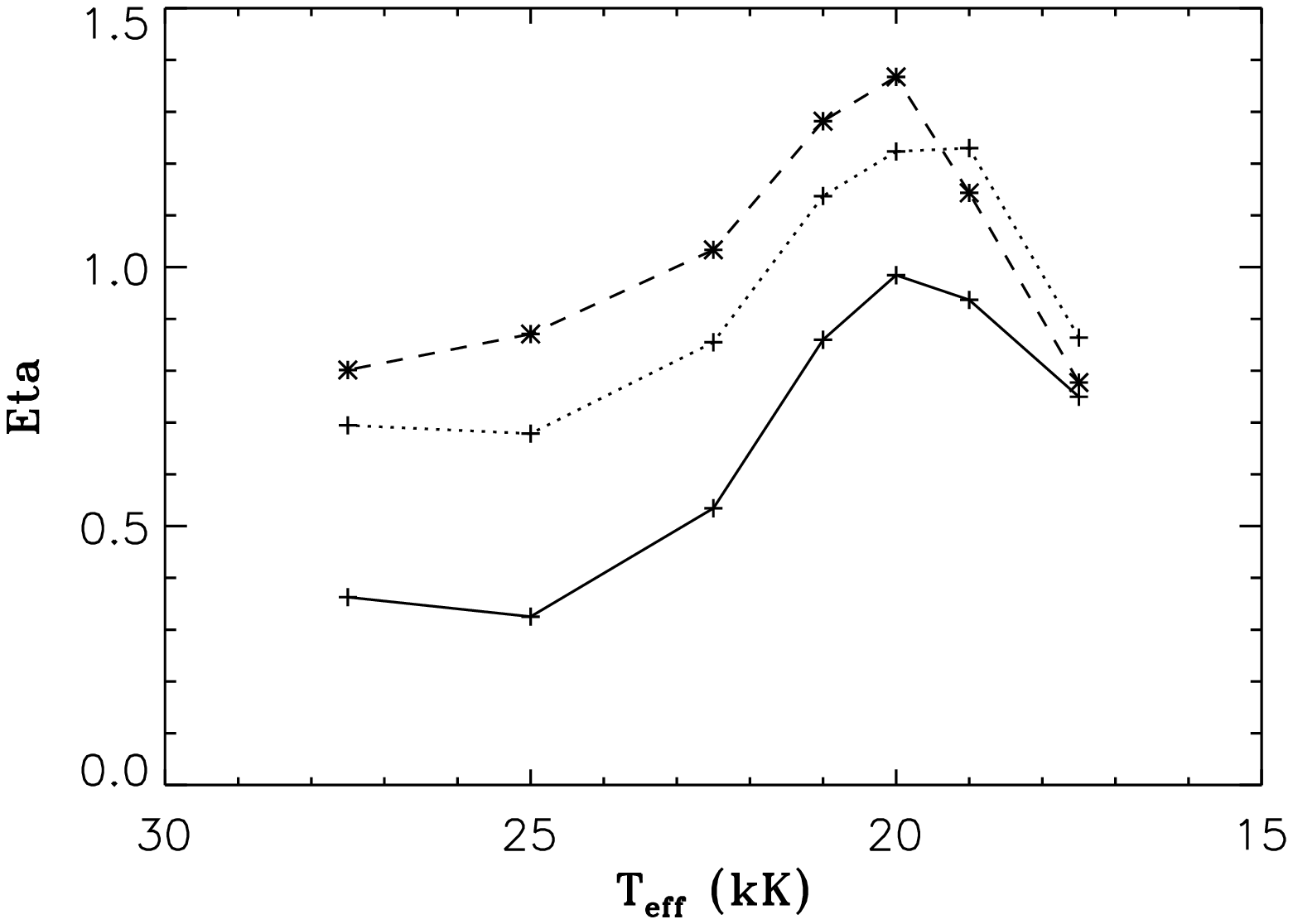, width = 9 cm}}
\caption{The predicted wind efficiency number $\eta$ versus \teff\ for the 3 different LBV mass models. The
solid line is for $M = 35\,\msun$, the dotted line for $M = 25\,\msun$, whilst the dashed line is for $M = 23\,\msun$,
representing $\Gamma$ values of respectively 0.5, 0.7 and 0.8.}
\label{f_lbv-eta}
\end{figure}

We adopt similar hydrogen (X=0.38) and helium (Y=0.60)
fractions as we adopted in Vink \& de Koter (2002), and as a representative 
model set we chose a similar set of mass, luminosity, and associated Eddington $\Gamma$ 
factors as in that paper. For the first model set in Table 2 we chose a much lower
mass (of 35\,\msun) for the same luminosity of $\logl = 6.0$ as in the previous 
section, resulting in an Eddington factor of 0.5, 
whilst the next series of models were chosen to have masses of 25 and 23 solar masses -- for the 
same fixed luminosity -- resulting in Eddington factors of 0.7 and 0.8 respectively. 
Note that these Eddington $\Gamma$ values only include the opacity of electron scattering, and 
for a discussion on the total opacity, we refer the reader to Vink et al. (2011). 
The relevance of the Eddington parameter for mass-loss rates was highlighted for VMS 
in the VLT Flames Tarantula Survey (Bestenlehner et al. 2014).

\begin{table*}
\centering
\begin{tabular}{lllll|l|ccc}
\hline
\hline\\[-6pt]
\mstar  & $\log L$ & $\Gamma$ & \teff & $Z/\zsun$ & \vesc    & \vinf   & $\log \mdot$ & \(\beta\) \\[2pt]
[\msun] & [\lsun] &           & (kK) &           & [\kmsec] & [\kmsec] & [\msunyr]   &         \\[3pt]
\hline\\[-7pt]
35      & 6.0     & 0.5       & 27.5 & 1.0       & 551      & 1811    & $-$5.39      & 1.29\\ 
        &         &           & 25   &           & 501      & 1863    & $-$5.45      & 1.44\\
        &         &           & 22.5 &           & 451      & 1645    & $-$5.18      & 1.22\\
        &         &           & 21   &           & 421      & 961     & $-$4.74      & 0.97\\
        &         &           & 20   &           & 401      & 663     & $-$4.52      & 0.84\\
        &         &           & 19   &           & 380      & 457     & $-$4.38      & 0.76\\
        &         &           & 17.5 &           & 350      & 253     & $-$4.22      & 0.69\\
\hline
25      & 6.0     & 0.7       & 27.5 & 1.0       & 465      & 1259    & $-$4.95      & 1.43\\ 
        &         &           & 25   &           & 423      & 1412    & $-$5.01      & 1.68\\
        &         &           & 22.5 &           & 381      & 1289    & $-$4.87      & 1.46\\
        &         &           & 21   &           & 355      & 802     & $-$4.54      & 1.09\\
        &         &           & 20   &           & 339      & 557     & $-$4.35      & 0.93\\
        &         &           & 19   &           & 322      & 370     & $-$4.17      & 0.82\\
        &         &           & 17.5 &           & 296      & 237     & $-$4.13      & 0.70\\
\hline
23      & 6.0     & 0.8       & 27.5 & 1.0       & 446      & 1102    & $-$4.83      & 1.53\\ 
        &         &           & 25   &           & 406      & 1313    & $-$4.87      & 1.87\\
        &         &           & 22.5 &           & 365      & 1155    & $-$4.74      & 1.49\\
        &         &           & 21   &           & 341      & 718     & $-$4.44      & 1.10\\
        &         &           & 20   &           & 325      & 506     & $-$4.26      & 0.96\\
        &         &           & 19   &           & 308      & 321     & $-$4.14      & 0.82\\
        &         &           & 17.5 &           & 284      & 199     & $-$4.10      & 0.70\\
\hline
\end{tabular}
\caption{Wind models for a range of 3 series of LBV models, the first two of these are identical to 
those presented in Vink \& de Koter (2002) for comparison, whilst the third model series with $M = 23 \msun$ 
was added for its even higher Eddington $\Gamma$ parameter of 0.8.}
\label{tab:lbv}
\end{table*}

The results from Table 2 are plotted in Figs. (5) - (9), showing the mass-loss rates, wind 
terminal velocities, and wind velocity over escape velocity ratio, the wind structure parameter $\beta$, and 
wind efficiency number $\eta$, respectively. The 
results are qualitatively similar to the earlier 60 \msun\ supergiant model. Note that the 
jumps for these LBV models are steeper than for the supergiant model, and that in contrast 
to previous globally consistent predictions in Vink \& de Koter (2002) the jump for LBVs has 
now shifted to the correct effective temperature of approximately 21\,000 K. The reason
for the onset of the bi-stability jump at lower temperature than previously is thought to be 
that for the stars on the hot side of the jump the winds are now thinner (due to the lower mass-loss rates and 
higher wind velocities), and that the lower density causes the Fe {\sc iv} to {\sc iii} 
recombination of Vink et al. (1999) to occur at lower effective temperature.

An additional finding
is that we find the size of the bi-stability jump, perhaps unexpectedly, to be {\it less} pronounced 
at higher $\Gamma$ values. We interpret this to be the result of saturation. Higher $\Gamma$ models
already show higher mass-loss rates than lower $\Gamma$ models on the hot side of the bi-stability jump
(as well as associated lower terminal wind velocities), so there is less of an opportunity 
to increase dramatically on the cool side of the bi-stability jump.

The absolute values of the terminal wind velocities are now down to 200 km/s, which is in 
the correct range for S Doradus type LBVs (Vink 2012). These slow winds are consistent 
with LBVs  being the direct progenitors of Type IIb and IIn supernovae inferred from the
the slow outflows (Kotak \& Vink 2006; 
Trundle et al. 2008; Smith et al. 2008; Groh \& Vink 2011; Groh 2014; Gr\"afener \& Vink 2016), 
and inconsistent with the fast outflows expected from 
Wolf-Rayet stars (Soderberg et al. 2006; Gal-Yam et al. 2014).

Note that the steepest part of the increase in the mass-loss rate is seen in
the temperature range 22- 20 kK, i.e. at {\it lower} values of \teff\ than suggested in 
the Vink et al. 2000/2001 mass-loss recipe. These lower \teff\ values of the bi-stability are
in good accord with both the observed drop in terminal wind velocity around spectral type B1
(Lamers et al. 1995; Crowther et al. 2006), as well as {\sc cmfgen} modelling
by Petrov et al. (2014; 2016). This means that our earlier attribution of the bi-stability jump 
temperature offset to the use of the modified nebular approximation was not correct. The reason
for the discrepancy was the semi-empirical nature of the earlier approach instead.

Figure \ref{f_lbv-beta} shows the behaviour of the wind structure parameter $\beta$ as a function of temperature. 
Whilst values on both the cool and the hot side of the bi-stability jump are in the range 0.7 - 1.5, and 
in general accord with earlier CAK-type models (Pauldrach et al. 1986; M\"uller \& Vink 2008; 
Muijres et al. 2012; Krticka et al. 2016), {\it at} the bi-stability jump the $\beta$ parameter peaks 
at larger values of order 1.5 - 2. We attribute this to several ionization stages of the driving ions to be 
playing a role, and in more sophisticated modelling a $\beta$ law would not be appropriate 
to describe the dynamical wind behaviour in detail (see e.g. Sander et al. 2018). We wish to emphasize 
that for models on the cool side of the bi-stability jump below 20 kK the derived $\beta$ values are no larger
than for O star models, i.e. in the range 0.7 - 1. 
This contrasts with the high $\beta$ values of up to 2 - 3 derived from empirical H$\alpha$ modelling 
(Trundle et al. 2004; Crowther et al. 2006), which may be artificially large
due to the neglect of optically thick (macro) clumping in the atmosphere modelling (Petrov et al. 2014).

Finally, we note that we did not converge dynamical models at \teff\ values 
in the range of the {\it second} bi-stability jump around 10\,000 K (Lamers et al. 1995; Petrov et al. 2016).

\section{Metallicity Dependence of the bi-stability jump for LBVs}
\label{s_lbvz}
\begin{figure}
\centerline{\psfig{file=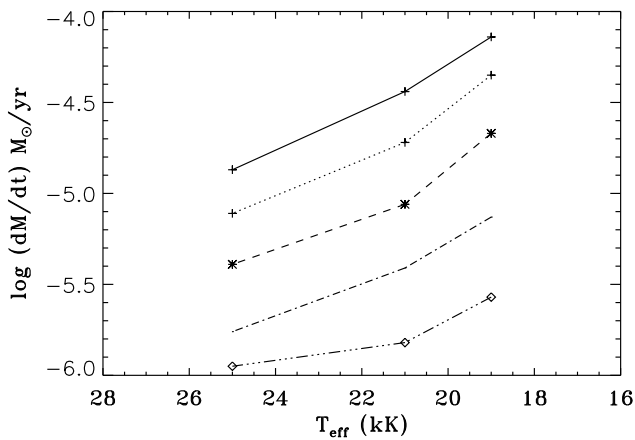, width = 9 cm}}
\caption{Predicted LBV mass-loss rates ($\mdot$) versus \Teff\ for five metallicities. 
The solid line is for solar metallicity, whilst the dotted, dashed, dotted-dashed, and dot-dot-dot-dashed lines 
are for a third, a tenth, 3.33\% and 1\% solar metallicity, respectively.}
\label{f_bist-Z-mdot}
\end{figure}

\begin{figure}
\centerline{\psfig{file=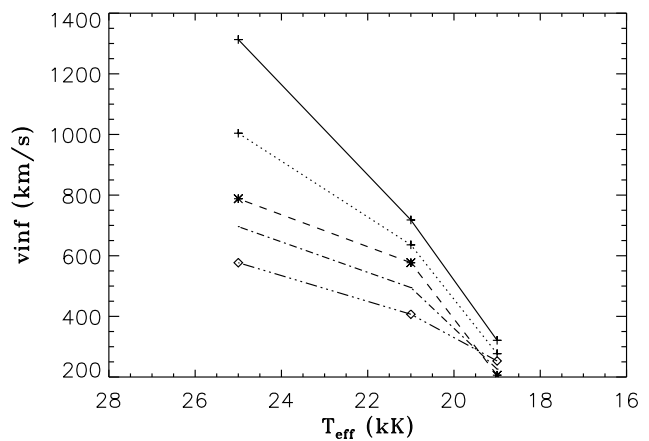, width = 9 cm}}
\caption{Predicted LBV wind velocities ($\vinf$) versus \teff\ for five metallicities. The symbols are the 
same as in Fig.\,\ref{f_bist-Z-mdot}}
\label{f_bist-Z-vinf}
\end{figure}

As a next step in our modelling we vary the metal contents $Z$, in order to investigate
if the size of the bi-stability jump is expected to be different in lower $Z$ galaxies, which are 
also representative of massive stars at
earlier Cosmic times. For this purpose we zoom in on the relevant \teff\ range for the bi-stability
jump over a $Z$ range varying from solar, to values as low as 1\% solar, with results listed in Table\,3 and 
plotted in Figs. 10 \& 11. 

\begin{table*}
\centering
\begin{tabular}{llll|ccc}
\hline
\hline\\[-6pt]
\mstar  & $\log L$    & $Z/\zsun$ & \teff & \vinf   & $\log \mdot$ & \(\beta\) \\[2pt]
[\msun] & [\lsun]     &           & (kK)  & [\kmsec] & [\msunyr]   &         \\[3pt]
\hline\\[-7pt]
23      & 6.0          & $1/3$    & 25    & 1004    & $-$5.11      & 1.53\\
        &              &          & 21    &  636    & $-$4.72      & 1.10\\           
        &              &          & 19    &  277    & $-$4.35      & 0.78\\          
\hline
23      & 6.0          & $1/10$   & 25    & 788     & $-$5.39      & 1.32\\
        &              &          & 21    & 577     & $-$5.06      & 1.07\\           
        &              &          & 19    & 206     & $-$4.67      & 0.70\\           
%        &              &          & 17    & 123     & $-$4.68      & 0.68\\          
\hline
23      & 6.0          & $1/33$   & 25    & 696    & $-$5.76       & 1.14\\
        &              &          & 21    & 495    & $-$5.41       & 1.00\\           
        &              &          & 19    & 226    & $-$5.13       & 0.76\\           
%        &              &          & 17    & 197    & $-$5.18       & 0.84\\          
\hline
23      & 6.0          & $1/100$  & 25    & 577    & $-$5.95       & 0.92\\
        &              &          & 21    & 407    & $-$5.82       & 0.96\\   
        &              &          & 19    & 253    & $-$5.57       & 0.75\\           
%        &              &          & 17    & 208    & $-$5.59       & 0.81\\          
\hline
\end{tabular}
\caption{Wind models for the same 3 LBV mass ranges as shown before, but now as a function of $Z$.}
\label{tab:lbvZ}
\end{table*}

As expected mass-loss rates generally drop with lower $Z$ (Fig.\,10), whilst terminal
wind velocities display opposite behaviour with \teff\ (Fig.\,11). 
At \teff\ values below the bi-stability jump, terminal 
velocities seem to converge to similar values for all metallicities (see also Vink 2018).
The size of the bi-stability jump does indeed appear to be a function of $Z$, with
larger $Z$ giving rise to a larger bi-stability jump, due to an increase in the role
of Fe in the line driving at higher $Z$ (Vink et al. 2001). 

Kalari et al. (2018) recently investigated the incidence of S Doradus variability amongst 
normal B supergiants in the low metallicity environment of the Small Magellanic Cloud (SMC), 
finding a surprisingly low number of S Dor variables in the SMC. This may be related
to our finding of lower $Z$ leading to a smaller bi-stability jump.

\section{Summary}
\label{s_sum}

We presented mass-loss predictions from Monte Carlo radiative 
transfer models for early-type supergiants and LBVs, and we found that:

\begin{itemize}

\item{The previously discovered observed drop in terminal wind velocities at spectral type B1 is confirmed 
by our dynamically consistent supergiant models.}

\item{The bi-stability jump in mass-loss rate is stronger than was derived in previous 
Monte Carlo modelling.}

\item{This would imply that within the rotationally induced bi-stability model of Pelupessy et al. (2000) for B[e] supergiants, the 
expected density contrast between the hotter pole and cooler equator could increase by up to one order of magnitude -- to a factor 100 -- which 
may be sufficient to account for the disk densities of B[e] supergiants, although the disk velocity structure would still
need to be explained}.

\item{Our wind predictions may have relevance for the slow wind inferred for the HMXB IGR J17252-3616, or other 
HMXBs.}
 
\item{The temperature of the bi-stability jump is now at the observed location of 21\,000 K, in agreement with {\sc cmfgen} models. 
This boosts confidence in the applicability of the modified nebular approximation.}

\item{The bi-stability jump is larger at {\it lower} Eddington $\Gamma$ parameter.}

\item{The bi-stability jump is larger at higher metallicity.}

\end{itemize}

%*********************************************************************

\begin{acknowledgements}

I would like to thank the anonymous referee for a constructive report. 
And I acknowledge hospitality of the Kavli Institute for Theoretical
Physics (KITP), Santa Barbara, and to Ed van den Heuvel for introducing me to 
the problem of the slow wind in IGR J17252-3616 during my stay at KITP, which was 
supported by the National
Science Foundation under Grant No. NSF PHY11-25915

\end{acknowledgements}

\end{document}